\documentclass[%
 reprint,
 amsmath,amssymb,
 aps,
]{revtex4-2}

\usepackage{graphicx}
\usepackage{dcolumn}
\usepackage{bm}

\usepackage[usenames]{color}
\usepackage{colortbl}
\begin{document}

\preprint{APS/123-QED}

\title{Controlling the water nonlinear refractive index in the THz frequency range \\ via temperature variation}

\author{Aleksandra Nabilkova}
\author{Azat Ismagilov}

  \author{Maksim Melnik}
 \author{Anton Tcypkin}
 \email{tsypkinan@itmo.ru}

 \affiliation{%
International Laboratory of Femtosecond Optics and Femtotechnologies, ITMO University, St. Petersburg, Russia. \\
Laboratory of Quantum Process and Measurements, ITMO University, St. Petersburg, Russia.
}%

 \author{Mikhail Guselnikov}
 \author{Sergei Kozlov}
 \affiliation{%
International Laboratory of Femtosecond Optics and Femtotechnologies, ITMO University, St. Petersburg, Russia.
}%

\author{Xi-Cheng Zhang}
\affiliation{
The Institute of Optics, University of Rochester, Rochester, NY, USA}%

\date{\today}

\begin{abstract}
To create self-controlled radiation photonics systems, it is necessary to have complete information about the nonlinear properties of the materials used. In this paper, the vibrational mechanism of the giant low–inertia cubic nonlinearity of water in the terahertz frequency range is experimentally proven. Its dominance which manifests itself when the temperature of the liquid changes is demonstrated. The measured nonlinear refractive index in the THz frequency range for water jet at temperatures from 14$^{\circ}$C to 21$^{\circ}$C demonstrates a correlation with the theoretical approach, varies from 4 to 10$\cdot$10$^{-10}$ cm$^2$/W and is characterized by a inertial time constant of less than 1 ps. 
\end{abstract}

\maketitle


\section{Introduction}

In recent years the area of nonlinear terahertz (THz) photonics develops intensively. This process results from recent investigation of high-intensity THz radiation sources \cite{Hafez_2016}, which might show applicability in communications \cite{KoenigComun}, non-destructive evaluation \cite{s20030712,inproceedingsZimdars}, light-control devices etc. \cite{2019cant.book, articleKamp, SALEN20191}. 

THz sources of ultrafast pulses gave an opportunity to construct devices which can control radiation parameters due to nonlinear effects of pulses self-modulation \cite{articleTurchi, articleWang, ZengGongWangZhouZhangLanCongWangSongZhaoYangMittleman+2022+415+437}. However, there is still a problem with gaining high intensities, which is essential to observe nonlinear effects. 

In 2015 authors of \cite{dolgaleva2015prediction} supposed that dominant mechanism of nonlinearity in THz spectral range is based on anharmonic oscillations of atoms which the molecules of matter consist of. That theory predicted a giant nonlinear refractive index of some liquid and crystal materials in THz range \cite{dolgaleva2015prediction, zhukova2020estimations}. Later, various scientific groups reported experimentally obtained giant value of nonlinear refractive index in liquid water and some other liquids and crystals via different methods (e.g. z-scan, full-phase analysis) \cite{tcypkin2019high, novelli2020nonlinear, francis2020terahertz, tcypkin2021giant, novelli2022terahertz}. These discoveries proved predictions for the theory of vibrational nonlinearity in THz range and opened wide perspectives for using nonlinear effects in ultrafast THz photonics. 

The analytical formula for nonlinear refractive index contains thermal expansion coefficient in power of two, which depends on temperature. In the case of liquid water this coefficient equals to zero near 4$^{\circ}$C, which must result in zero nonlinear refractive index at this temperature. The authors of \cite{novelli2022temperature} reported that the nonlinear response of liquid water at 1 THz is similar at 21$^{\circ}$C and 4$^{\circ}$C and, based on that result, declared that the theory of vibrational nonlinearity in THz range is the wrong one. However, it is worth noting that there is no reason to estimate the nonlinear response of the refractive index in the THz frequency range based on nonlinear absorption. Direct contribution to the nonlinear refractive index in THz frequency range comes from the resonant interaction of oscillations in the IR range (100 THz = 3 $\mu$m), and not in the THz. To experimentally demonstrate this statement, we conducted a series of experiments on temperature dependency of liquid water nonlinear refractive index measurements and compare it with the curve calculated via theory of vibrational nonlinearity. The measurements were provided by z-scan method with central frequency of single-cycle pulse at 0.75 THz. The results obtained show the possibility of controlling the nonlinear properties of water by changing its temperature and confirm the vibrational nature of the nonlinearity in the THz frequency range. This can be used to create devices for ultrafast THz photonics.

\section{Experimental setup}

\begin{figure}[h!]
\centering\includegraphics[width=0.9\columnwidth]{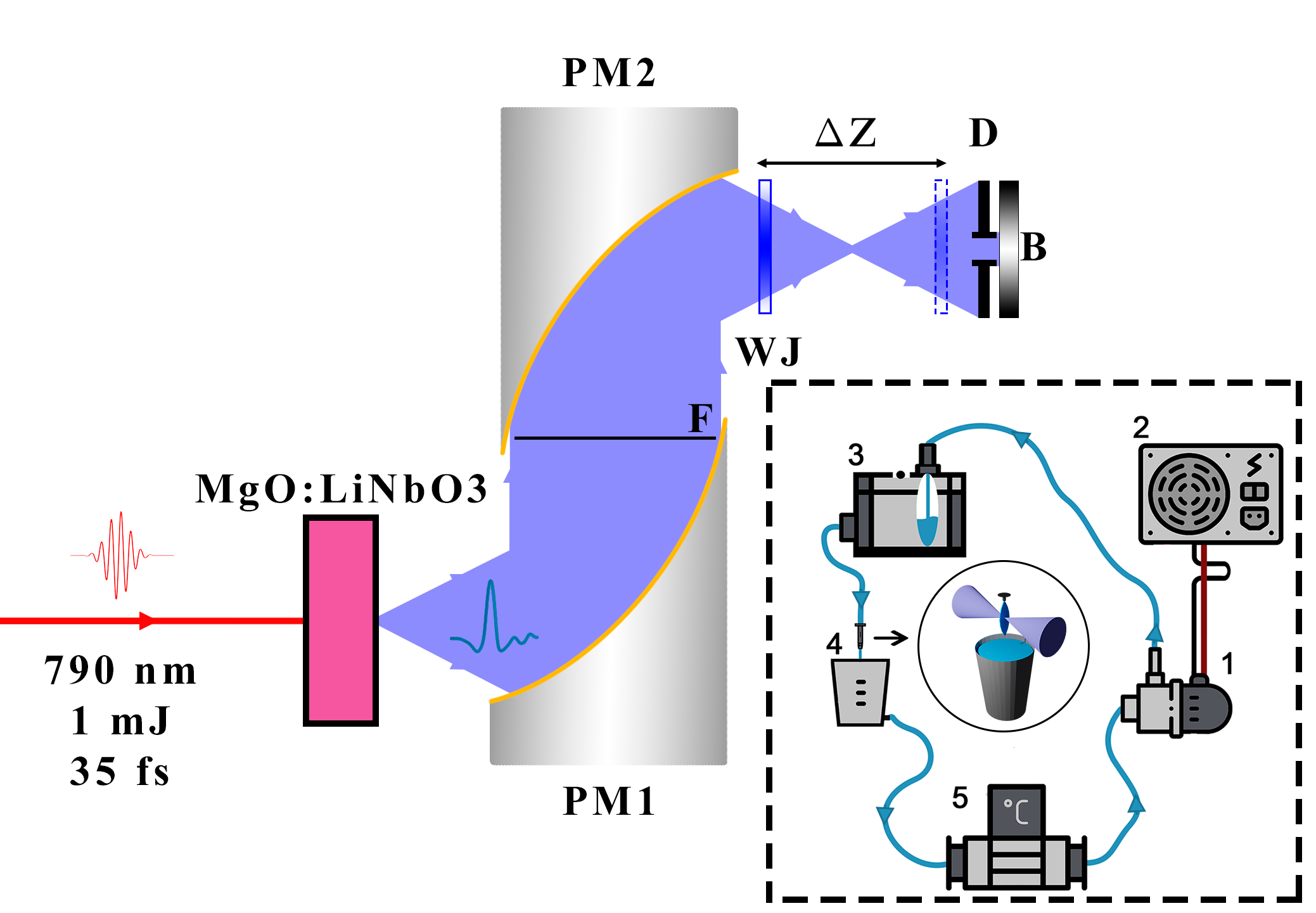}
\caption{ Optical scheme of the experimental setup: THz beam, after its generation in MgO:LiNbO$_3$ crystal, is collimated by the parabolic mirror (PM1) and directed to the second parabolic mirror (PM2) on which it is focused; teflon filter (F) is used to filter pump radiation; water jet (WJ) moves along the radiation propagation axis $\Delta Z$ before bolometer (B) diaphragm (D) is placed, which make it possible to use the method of open and closed aperture. The insertion shows scheme of water jet formation system: 1 is pump, 2 is power supply, 3 is accumulator tank, 4 is nozzle, 5 is temperature control unit.}
\label{fig1}
\end{figure}

The nonlinear refractive index n$_2$ of water temperature dependence verification consisted of a series of experiments based on the classical z-scan method \cite{Sheik_Bahae_1990} using a pulsed THz radiation source based on a lithium niobate crystal (see Supplemental material \cite{Supplemental}, Section A for details). The experimental setup is shown in Fig. \ref{fig1}. The generation of a high-intensity THz field is carried out due to the effect of optical rectification in a MgO:LiNbO$_3$ crystal \cite{F_l_p_2012} of a femtosecond optical pulse with the following parameters: pulse energy 1 mJ, pulse duration 35 fs, center wavelength 790 nm and repetition rate 1 kHz. The intensity of THz radiation when focused by a parabolic mirror with a focal length of 1 inch is 10$^8$ W/cm$^2$.

A distinctive feature of the experiment was the use of a liquid jet \cite{Watanabe_1989} rather than a cell, which, as shown in our previous works \cite{tcypkin2019high, tcypkin2021giant}, makes it possible to avoid the cumulative thermal effect that can affect the nonlinearity. The water temperature was controlled by a cooling unit in a jet formation system similar to that presented in \cite{ismagilov2021liquid}, the temperature varying from 14$^{\circ}$C to 24$^{\circ}$C. The sensitivity of evaluation method and features of cooling system do not permit operating on temperatures below 14 $^{\circ}$C due to an increase in measurement error. To move along the radiation propagation axis, the nozzle was mounted on a linear translator. The measurement of the THz field distribution was carried out using the method of open and closed aperture \cite{Sheik_Bahae_1990}, in order to exclude the influence of nonlinear absorption on the results.

\section{Results and discussion}

\begin{figure}[h!]
\centering\includegraphics[width=\columnwidth]{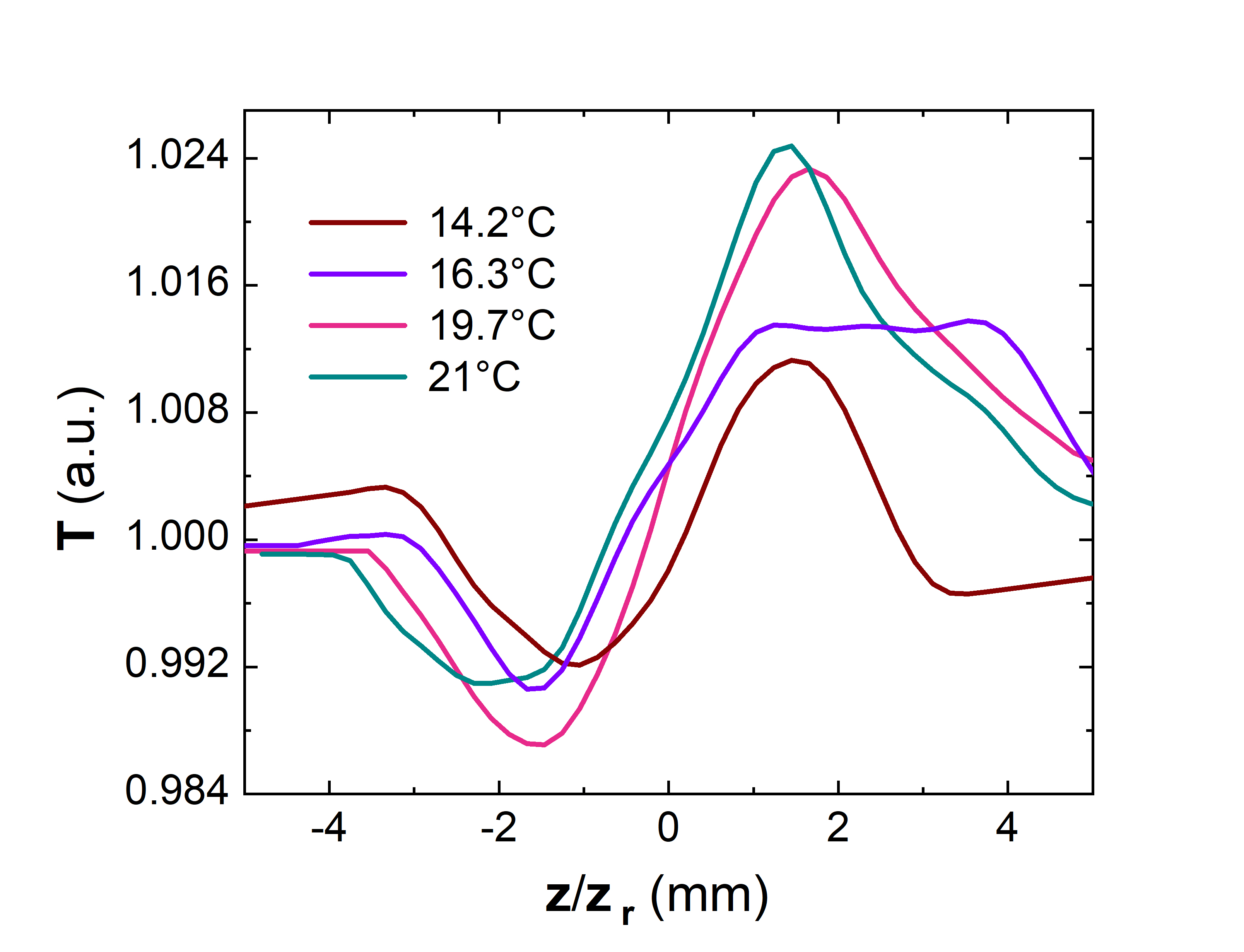}
\caption{Experimental z-scan curves of water with different temperatures.}
\label{fig2}
\end{figure}

Figure \ref{fig2} shows the resulting z-scan curves obtained by dividing the data from a closed aperture case by the data with an open aperture case. These curves were normalized to the THz pulse energy in the linear propagation mode. The jet displacement range along the radiation propagation axis was 10 Rayleigh lengths. The Rayleigh length with the focusing parameters used is 0.96 mm.

As can be seen from Fig.\ref{fig2}, the peak-to-valley ratio changes drastically with water temperature. The higher it is, the greater the difference. Using the standard formula for calculating n$_2$ from the the ratio \cite{Sheik_Bahae_1990}, while satisfying the recommendations for working with few-cycle pulses \cite{Melnik_2019}, the values of n$_2$ for different temperatures were obtained. They are presented in Table \ref{tab1} (see Supplemental material \cite{Supplemental}, Section B for details).

\begin{table}[h!]
\begin{ruledtabular}
\caption{Nonlinear refractive index of water for different temperatures.}
\label{tab1}
\begin{tabular}{ccccccccc}
t$^{\circ}$C & 14 & 16 & 17 & 18 & 18.5 & 20 & 21  \\ 
n$_2$, 10$^{-10}$ cm$^2$/W & 4.2 & 5.3 & 6.5 & 6.8 & 7.8 & 9.5 & 9.7
\end{tabular}
\end{ruledtabular}
\end{table}

Using the mathematical model of the non-resonant vibrational contribution to the nonlinear refractive index \cite{dolgaleva2015prediction} (see Supplemental material \cite{Supplemental}, Section C for details), the dependence of n$_2$(t) on temperature can be written as follows:

\begin{eqnarray}
    n^{\omega\ll\omega_0}_2 (t) = \frac{{3a^2_l}m^2{\omega^4_0}{\alpha_T(t)^2}}{32n_0\pi^2q^2N^2{k^2_B}}
    \left[\left(n^{\omega\ll\omega_0}_{0,\nu} \right)^2 - 1\right]^3 \nonumber\\
    - \frac{9}{32{\pi}N{n_0}\hbar{\omega_0}}
    \left[\left(n^{\omega\ll\omega_0}_{0,\nu} \right)^2 - 1 \right]^2
    \label{eq1}
\end{eqnarray}

It can be seen that the only term in the expression (\ref{eq1}) that depends on temperature is the thermal expansion coefficient $\alpha_T(t)$, the values of which for water were obtained in \cite{articleKell}.

\begin{figure}[h!]
\centering\includegraphics[width=\columnwidth]{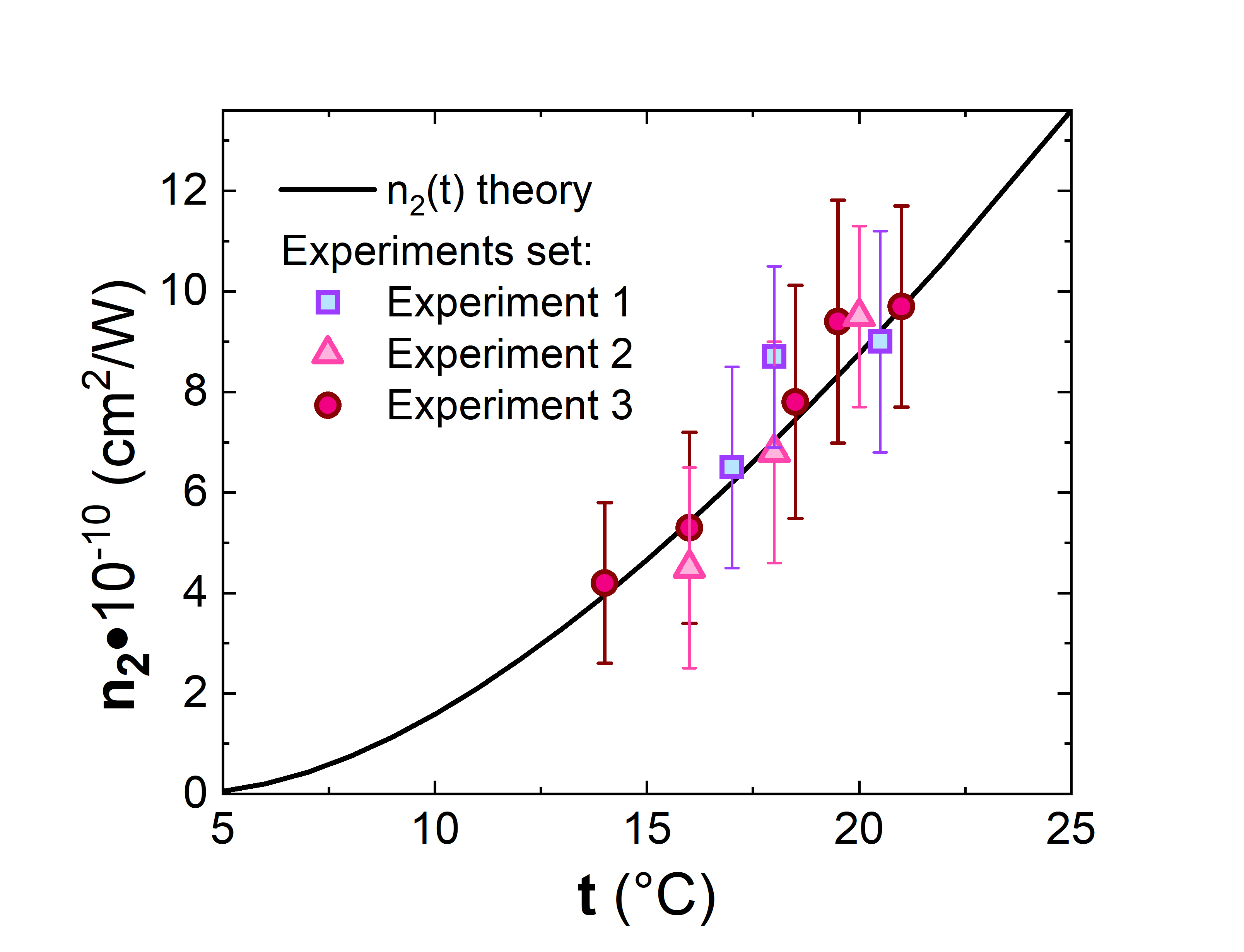}
\caption{Nonlinear refractive index n$_2$ dependence on water temperature based on theoretical model \cite{dolgaleva2015prediction} (black solid curve). Nonlinear refractive index n$_2$ of water for different temperatures from experimental z-scan measurements (colored dots).}
\label{fig3}
\end{figure}

Figure \ref{fig3} shows the dependence of n$_2$ on temperature for water, calculated by the formula (\ref{eq1}) according to the values of the thermal expansion coefficient $\alpha_T(t)$ and superimposed on experimental data.

Experimental data sets were taken independently from each other to confirm the repeatability. The results obtained demonstrate that a decrease in water temperature leads to a decrease in the value of n$_2$, which is in full agreement with the presented theoretical model. This confirms that the resulting cubic nonlinearity is of an vibrational nature.

\section{Conclusion}

In the article, the temperature dependence of the nonlinear refractive index of water in the THz spectral range has been measured for the first time. It has been experimentally shown that, when the temperature changes from 14$^{\circ}$C to 21$^{\circ}$C, the nonlinear refractive index value of water in spectral range from 0.2 to 1 THz changes from 4$\cdot$10$^{-10}$ cm$^2$/W to 10$\cdot$10$^{-10}$ cm$^2$/W. Influence of cumulative thermal effect on the nonlinearity was avoided by the use of a liquid jet instead of a cell. Since the measurements of the nonlinear characteristic of water were carried out using the jet, each subsequent THz pulse interacted with a new region of the water jet, therefore, the inertia of the nonlinearity mechanism did not exceed 1 ps. It is theoretically shown that the measured dependence of the water nonlinear refractive index on temperature corresponds to the dependence for the squared coefficient of thermal expansion of water on temperature. This is a new and significant confirmation of the vibrational nature of the refractive index giant low-inertia nonlinearity of liquids in the THz range. This confirmation is of great importance for future development of devices for ultrafast THz photonics based on materials with such a high nonlinearity.

\begin{acknowledgments}
The study is funded by the Ministry of Science and Higher Education of the Russian Federation (Passport No. 2019-0903).
\end{acknowledgments}

\bibliographystyle{ieeetr}
\bibliography{apssamp}

\begin{thebibliography}{10}

\bibitem{Hafez_2016}
H.~A. Hafez, X.~Chai, A.~Ibrahim, S.~Mondal, D.~F{\'{e}}rachou, X.~Ropagnol,
  and T.~Ozaki, ``Intense terahertz radiation and their applications,'' {\em
  Journal of Optics}, vol.~18, p.~093004, aug 2016.

\bibitem{KoenigComun}
S.~Koenig, D.~Lopez-Diaz, J.~Antes, J.~Antes, F.~Boes, F.~Boes, R.~Henneberger,
  A.~Leuther, A.~Tessmann, R.~Schmogrow, R.~Schmogrow, D.~Hillerkuss,
  D.~Hillerkuss, R.~Palmer, T.~Zwick, C.~Koos, W.~Freude, O.~Ambacher,
  J.~Leuthold, J.~Leuthold, I.~Kallfass, and I.~Kallfass, ``{Wireless sub-THz
  communication system with high data rate },'' {\em Nature Photonics}, vol.~7,
  no.~12, pp.~977--981, 2013.

\bibitem{s20030712}
Y.~H. Tao, A.~J. Fitzgerald, and V.~P. Wallace, ``Non-contact, non-destructive
  testing in various industrial sectors with terahertz technology,'' {\em
  Sensors}, vol.~20, no.~3, 2020.

\bibitem{inproceedingsZimdars}
D.~Zimdars, J.~White, G.~Stuk, A.~Chernovsky, G.~Fichter, and S.~Williamson,
  ``Large area terahertz imaging and non-destructive evaluation applications,''
  vol.~48, 07 2006.

\bibitem{2019cant.book}
T.~{Elsaesser}, K.~{Reimann}, and M.~{Woerner}, {\em {Concepts and Applications
  of Nonlinear Terahertz Spectroscopy}}.
\newblock 2019.

\bibitem{articleKamp}
T.~Kampfrath, K.~Tanaka, and K.~Nelson, ``Resonant and nonresonant control over
  matter and light by intense terahertz transients,'' {\em Nature Photonics},
  vol.~7, pp.~680--690, 08 2013.

\bibitem{SALEN20191}
P.~Salén, M.~Basini, S.~Bonetti, J.~Hebling, M.~Krasilnikov, A.~Y. Nikitin,
  G.~Shamuilov, Z.~Tibai, V.~Zhaunerchyk, and V.~Goryashko, ``{ Matter
  manipulation with extreme terahertz light: Progress in the enabling THz
  technology},'' {\em Physics Reports}, vol.~836-837, pp.~1--74, 2019.

\bibitem{articleTurchi}
D.~Turchinovich, J.~Hvam, and M.~Hoffmann, ``Self-phase modulation of a
  single-cycle terahertz pulse by nonlinear free-carrier response in a
  semiconductor,'' {\em Physical Review B}, vol.~85, p.~201304, 02 2012.

\bibitem{articleWang}
Z.~Wang, J.~Qiao, S.~Zhao, S.~Wang, C.~He, X.~Tao, and S.~Wang, ``Recent
  progress in terahertz modulation using photonic structures based on
  two‐dimensional materials,'' {\em InfoMat}, vol.~3, 09 2021.

\bibitem{ZengGongWangZhouZhangLanCongWangSongZhaoYangMittleman+2022+415+437}
H.~Zeng, S.~Gong, L.~Wang, T.~Zhou, Y.~Zhang, F.~Lan, X.~Cong, L.~Wang,
  T.~Song, Y.~Zhao, Z.~Yang, and D.~M. Mittleman, ``A review of terahertz phase
  modulation from free space to guided wave integrated devices,'' {\em
  Nanophotonics}, vol.~11, no.~3, pp.~415--437, 2022.

\bibitem{dolgaleva2015prediction}
K.~Dolgaleva, D.~V. Materikina, R.~W. Boyd, and S.~A. Kozlov, ``{Prediction of
  an extremely large nonlinear refractive index for crystals at terahertz
  frequencies},'' {\em Physical Review A}, vol.~92, no.~2, p.~023809, 2015.

\bibitem{zhukova2020estimations}
M.~Zhukova, M.~Melnik, I.~Vorontsova, A.~Tcypkin, and S.~Kozlov, ``{Estimations
  of low-inertia cubic nonlinearity featured by electro-optical crystals in the
  THz range},'' vol.~7, no.~4, p.~98, 2020.

\bibitem{tcypkin2019high}
A.~N. Tcypkin, M.~V. Melnik, M.~O. Zhukova, I.~O. Vorontsova, S.~E. Putilin,
  S.~A. Kozlov, and X.-C. Zhang, ``{High Kerr nonlinearity of water in THz
  spectral range},'' {\em Optics express}, vol.~27, no.~8, pp.~10419--10425,
  2019.

\bibitem{novelli2020nonlinear}
F.~Novelli, C.~Y. Ma, N.~Adhlakha, E.~M. Adams, T.~Ockelmann, D.~Das~Mahanta,
  P.~Di~Pietro, A.~Perucchi, and M.~Havenith, ``{Nonlinear terahertz
  transmission by liquid water at 1 THz},'' {\em Applied Sciences}, vol.~10,
  no.~15, p.~5290, 2020.

\bibitem{francis2020terahertz}
K.~J.~G. Francis, M.~L.~P. Chong, E.~Yiwen, and X.-C. Zhang, ``Terahertz
  nonlinear index extraction via full-phase analysis,'' {\em Optics Letters},
  vol.~45, no.~20, pp.~5628--5631, 2020.

\bibitem{tcypkin2021giant}
A.~Tcypkin, M.~Zhukova, M.~Melnik, I.~Vorontsova, M.~Kulya, S.~Putilin,
  S.~Kozlov, S.~Choudhary, and R.~W. Boyd, ``Giant third-order nonlinear
  response of liquids at terahertz frequencies,'' {\em Physical Review
  Applied}, vol.~15, no.~5, p.~054009, 2021.

\bibitem{novelli2022terahertz}
F.~Novelli, C.~Hoberg, E.~M. Adams, J.~M. Klopf, and M.~Havenith, ``{Terahertz
  pump--probe of liquid water at 12.3 THz},'' {\em Physical Chemistry Chemical
  Physics}, vol.~24, no.~2, pp.~653--665, 2022.

\bibitem{novelli2022temperature}
F.~Novelli, C.~Millon, J.~Schmidt, S.~Ramos, E.~P. van Dam, A.~Buchmann,
  C.~Saraceno, and M.~Havenith, ``Temperature-independent non-linear terahertz
  transmission by liquid water,'' {\em arXiv preprint arXiv:2206.03998}, 2022.

\bibitem{Sheik_Bahae_1990}
M.~Sheik-Bahae, A.~Said, T.-H. Wei, D.~Hagan, and E.~V. Stryland, ``Sensitive
  measurement of optical nonlinearities using a single beam,'' {\em {IEEE}
  Journal of Quantum Electronics}, vol.~26, pp.~760--769, apr 1990.

\bibitem{Supplemental}
``See supplemental material for further details and derivations.'' \url{}.

\bibitem{F_l_p_2012}
J.~A. Fülöp, L.~P{\'{a}}lfalvi, S.~Klingebiel, G.~Alm{\'{a}}si, F.~Krausz,
  S.~Karsch, and J.~Hebling, ``Generation of sub-{mJ} terahertz pulses by
  optical rectification,'' {\em Optics Letters}, vol.~37, p.~557, feb 2012.

\bibitem{Watanabe_1989}
A.~Watanabe, H.~Saito, Y.~Ishida, M.~Nakamoto, and T.~Yajima, ``A new nozzle
  producing ultrathin liquid sheets for femtosecond pulse dye lasers,'' {\em
  Optics Communications}, vol.~71, pp.~301--304, jun 1989.

\bibitem{ismagilov2021liquid}
A.~O. Ismagilov, E.~A. Ponomareva, M.~O. Zhukova, S.~E. Putilin, B.~A.
  Nasedkin, and A.~N. Tcypkin, ``Liquid jet-based broadband terahertz radiation
  source,'' {\em Optical Engineering}, vol.~60, no.~8, p.~082009, 2021.

\bibitem{Melnik_2019}
M.~Melnik, I.~Vorontsova, S.~Putilin, A.~Tcypkin, and S.~Kozlov, ``Methodical
  inaccuracy of the z-scan method for few-cycle terahertz pulses,'' {\em
  Scientific Reports}, vol.~9, jun 2019.

\bibitem{articleKell}
G.~S. Kell, ``Precise representation of volume properties of water at one
  atmosphere,'' {\em J. Chem. Eng. Data}, vol.~12, p.~66–69, 01 1967.

\end{thebibliography}

\end{document}